\documentclass[aps,twocolumn,letterpaper, nofootinbib, floatfix, superscriptaddress,amsmath,amssymb]{revtex4-2}
\usepackage{amsmath,amssymb, graphicx}
\usepackage{bm,times, color}
\usepackage[utf8]{inputenc}
\usepackage[colorlinks,breaklinks]{hyperref}
\usepackage[svgnames]{xcolor}
\hypersetup{linkcolor=DarkBlue, citecolor=DarkBlue, filecolor=black, urlcolor=DarkBlue}
\usepackage{comment, mathrsfs,slashed}
\usepackage{amsfonts}
\usepackage{bbm}
\usepackage[normalem]{ulem}
\def \p{\partial}

\def \mb{\mathbf}

\def \ga{\gamma}

    \makeatletter
\def\@fnsymbol#1{\ensuremath{\ifcase#1\or \dagger\or \ddagger\or
   \mathsection\or \mathparagraph\or \|\or **\or \dagger\dagger
   \or \ddagger\ddagger \else\@ctrerr\fi}}
    \makeatother

\begin{document}
\title{Geometrically constrained particle dynamics revisited: Equation of motion in terms of the normal curvature of the constraint manifold}

\author{Wei-Han Hsiao}
\altaffiliation{Ph.D. in Physics, The University of Chicago}
\noaffiliation{}
\date{\today}

\begin{abstract}
We revisit the problem of the particle dynamics subject to a geometric holonomic constraint of codimension 1 in spatial dimensions $d = $2 and 3. In the absence of dissipation, we show that by solving the Lagrangian multiplier in a general fashion, the external potential independent part, the net normal force, of the equation of motion corresponds to precisely to the curvature of the trajectory on the constraint space multiplied by twice the kinetic energy. The tangent the trajectory is the instantaneous velocity. In $d=3$, this term equals the second fundamental form $\mathrm{I\!I}$ of the constraint surface evaluated on the unit tangent vector in the direction of velocity. Using these result we establish the relation between constrained particle dynamics with geodesic equations and derive intriguing kinematic implications using theorems from fundamental differential geometry.
\end{abstract}

\maketitle
\section{Introduction}
Ancient studies on differential geometry of curves is tangled with classical particle kinematics. Given a proper parametrization, a curve can be thought of as the trajectory of a particle under a prescribed motion. This picture in turn establishes one of the early bridges connecting geometry and physics. For instance, curvature, the reciprocal of the radius of the osculating circle, is proportional to the acceleration normal to the velocity. The concept of curvature, encompassing the planer one and the abstract high-dimensional generalizations, has become a critical object in the realm of physics.

From the preceding example, a mirrored question can be asked by swapping the prescribed object. Suppose we release a particle in a potential profile and impose a constraint that it must move in a designated subspace of the configuration space. What geometric information of the subspace shall be encoded in the equation of motion, which determines the trajectory?

More precisely, we consider the following class of problems: The particle dynamics is modeled by the standard non-relativistic Lagrangian in $\mathbb R^n$, $n=2$ or $3$:
\begin{align}
L = \frac{1}{2}m\dot{\mb x}^2 - V(\mb x)
\end{align}
subject to a time-independent holonomic constraint $f(\mb x) = 0$. We assume the solutions $\{\mb x\in\mathbb R^n| f(\mb x) = 0\}$ form a smooth manifold of dimension $n-1$. They define planar curves and smooth surfaces for $n = 2$ and $n = 3$ respectively.

There are a couple of procedures to tackle these problems in the conventional curriculum \cite{thornton2004classical, goldstein:mechanics}. If the constraint is solvable, a substitution and reduction of variables can render the original problem unconstrained. Another standard approach is to introduce a Lagrange multiplier to enforce the constraint. It enlarges the space of variables and the constraint becomes an instance of the Euler-Lagrange equations. Furthermore, it does not require guessing a clever coordinate and naturally introduces a normal force. The value of the Lagrange multiplier, nonetheless, is usually solved explicitly based on the specific system of interest, and its meaning remains inconspicuous. We note that in fact the value of the Lagrange multiplier can be solved for general smooth constraints \cite{dirac_1950, dirac2013lectures, weinberg_2015}. By virtue of the resulting equation of motion, we shall highlight the following facts:
\begin{enumerate}
\item The net normal force from the holonomic constraint is given by the normal curvature $\kappa_n$ of the constrained trajectory multiplied by twice the kinetic energy $mv^2$. In $d=2$, the subspace is a planar curve and $\kappa_n$ corresponds to the ordinary curvature $\kappa$. 
\begin{align}
(m\ddot{\mb x})_n = mv^2\kappa_n.
\end{align}
In $d=3$, the subspace is a surface and the normal force is the second fundamental form of the surface evaluated on the particle velocity $\mathrm{I\!I}(\dot{\mb x}, \dot{\mb x})$ multiplied by the mass. $\kappa_n$ is the curvature of trajectory the tangent vector of which is the particle velocity. 
\item In $d = 3$, in the absence of the external potential, the particle trajectory confined on the surface is a spatial geodesic, i.e., the curve of shortest length given end points on the surface. By spatial we mean it is different from the spacetime geodesic equation or a free fall in the context of general relativity.
\end{enumerate}
These results not only bind the normal force with a precise geometric property of the constraint surface rather its arbitrary data, but potentially bring new insights to the mature kinematics by utilizing the toolkits of differential geometry. A few kinematic implications from the global properties of curves, such as the rotation index and 4-vertex theorems, and local property of general embedded surfaces are made in the main text as a justification. 
 
The rest of the paper is organized as follows. In Sec.\ref{EoM}, we provide a thorough review of the constrained equation of motion for a rather general $f(\mb x)$. Using this result, we prove the above statements in Sec.~\ref{geometry} and~\ref{geodesic} using the idea from elementary differential geometry of curves and surfaces in $\mathbb R^3$. Some simple examples from classical mechanics are discussed using this new perspective. We derive several kinematic implications in Sec.\ref{applications} based on our results and some well-known theorems from geometry. Some remarks and outlook are compiled in Sec.\ref{conclusion}.
\section{Derivation of the equations of motion}\label{EoM}
We start off by reviewing the derivation of full equation of motion for a general smooth constraint. The Lagrangian derivation presented in this section substantially follows Ref \cite{weinberg_2015}. Note that in the 1st edition, there were a couple of assumptions that render the Weinberg's result incomprehensive , and the normal force term was missing in both Lagrangian and Hamiltonian formulations. We recently learnt it has been revised in the 2nd edition, which we recommend to the interested readers. The Hamiltonian approach mainly references the logic in Ref.\cite{dirac2013lectures, dirac_1950}.
\subsection{Lagrangian formulation}
We first derive the equation of motion with the Lagrangian formulation. Let the coordinates of the particle be $\mb x$ and the geometric holonomic constraint be $f(\mb x) = 0$. The extended Lagrangian to minimize is the sum of the ordinary particle Lagrangian and the product of the constraint and the corresponding Lagrange multiplier $\lambda$. 
\begin{align}
L = \frac{1}{2}m\dot{\mb x}^2 - V(\mb x) + \lambda f(\mb x).
\end{align}
The extended set of equations of motion are derived by varying $\mb x$ and $\lambda$:
\begin{subequations}
\begin{align}
& m\ddot{\mb x} = -\nabla V(\mb x) + \lambda\nabla f(\mb x)\\
& f(\mb x) = 0.
\end{align}
\end{subequations}
To solve for $\lambda$, it is helpful to consider the self-consistency conditions $d^nf/dt^n = 0$, $n = 1, 2$.
\begin{subequations}
\begin{align}
\label{secondary}& 0 = \frac{df}{dt} = \dot{\mb x}\cdot\nabla f = 0\\
\label{nonnew} & 0 = \frac{d^2f}{dt^2} = \ddot{\mb x}\cdot\nabla f + \dot{\mb x}\cdot\nabla\nabla f\cdot\dot{\mb x}.
\end{align}
\end{subequations}
The first condition entails the motion normal to the constraint surface $f(\mb x)$ should vanish. The second helps us to eliminate $\lambda$ through $\ddot{\mb x}$. To do this, let us first write Eq~\eqref{nonnew} $m\ddot{\mb x}\cdot\nabla f = -m\dot{\mb x}\nabla\nabla f\cdot\dot{\mb x}$ and project whole equation of motion onto $\nabla f$:
\begin{align*}
m\ddot{\mb x}\cdot\nabla f = -\nabla f\cdot\nabla V +\lambda|\nabla f|^2.
\end{align*}
Eliminating $\ddot{\mb x}$ yields 
\begin{align}
\lambda = \frac{1}{(\nabla f)^2}(\nabla f\cdot\nabla V - m\dot{\mb x}\cdot\nabla\nabla f\cdot\dot{\mb x}).
\end{align}
Consequently, the final equation of motion reads 
\begin{align}
\label{eof_L}m\ddot{\mb x} = -\left[1 - \frac{\nabla f\nabla f}{|\nabla f|^2}\right]\cdot\nabla V - \frac{m\nabla f}{|\nabla f|^2}(\dot{\mb x}\cdot\nabla\nabla f\cdot\dot{\mb x}).
\end{align}
The meaning of the decomposition is clear. The first term represents the effect of potential tangential to the surface $f$, and the second term is the effect of normal force confining the particle to the surface. Physically speaking, the second term must exist otherwise even the elementary planer circular motion cannot be explained in Newtonian mechanics. 

Besides, we note that the energy is still conserved regardless of $\lambda$. To see this fact explicitly, let us project~\eqref{eof_L} onto $\dot{\mb x}$ and integrate over $t$. Because of condition~\eqref{secondary}, the normal force is dropped by the projection and the integral from $t_0$ to $t_1$ yields:
\begin{align}
\frac{1}{2}m(\dot{\mb x}(t_1)^2 -\dot{\mb x}(t_0)^2   ) = -[V(\mb x(t_1)) - V(\mb x(t_0))].
\end{align}
\subsection{Hamiltonian formulation} 
The same equation of motion can be derived using Hamiltonian method. Since the Lagrangian has the trivial quadratic $\dot{\mb x}$ dependence, the standard component of Hamiltonian is given by $H = \mb p\cdot\dot{\mb x} -L$, where $\mb p = \frac{\p L}{\p \dot{\mb x}}$. To impose the constraint, it also requires a Lagrange multiplier $\lambda$ to form the total Hamiltonian $H_T$.
\begin{align}
H_T = \frac{\mb p^2}{2m} + V(\mb x) + \lambda f(\mb x) = H(\mb x, \mb p) + \lambda f(\mb x)
\end{align}
Different from the Lagrangian formulation, the time evolution is given by the action of Poisson bracket
\begin{subequations}
\begin{align}
& \dot{\mb x} = \{\mb x, H_T\}  = \{\mb x, H\} + \lambda\{\mb x, f\}\\
& \dot{\mb p}=\{\mb p, H_T\}  = \{\mb p, H\} + \lambda\{\mb p, f\},
\end{align}
\end{subequations}
where 
\begin{align}
\{ f, g\} = \frac{\p f}{\p \mb x}\cdot\frac{\p g}{\p \mb p} -\frac{\p f}{\p \mb p}\cdot\frac{\p g}{\p \mb x}.  
\end{align}
Elementary computation yields, 
\begin{subequations}
\begin{align}
\label{p_is_mv}& \dot{\mb x} = \mb p/m\\
& \dot{\mb p} = -\nabla V - \lambda \nabla f.
\end{align}
\end{subequations}
The constraint $f(\mb x) = 0$ shall satisfy the consistency conditions as follows. Note that in the right-hand side the constraint is imposed after evaluating the Poisson bracket. 
\begin{subequations}
\begin{align}
& \frac{df}{dt} = \{ f, H\} + \lambda \{ f, f\} = \{f, H\} = \nabla f\cdot\mb p = 0\\
& \frac{d}{dt}[\nabla f\cdot\mb p] = \{ \nabla f\cdot\mb p, H\} + \lambda\{ \nabla f\cdot\mb p, f\} = 0.
\end{align}
\end{subequations}
The last equation evaluates to 
\begin{align}
\frac{1}{m}\mb p\cdot\nabla\nabla f\cdot\mb p - \nabla f\cdot\nabla V = \lambda|\nabla f|^2.
\end{align}
Consequently, the equation for $\mb p$ reads
\begin{align}
\dot{\mb p} =& -\nabla V - \frac{\nabla f}{|\nabla f|^2}\left( \frac{1}{m}\mb p\cdot\nabla\nabla f\cdot\mb p - \nabla f\cdot\nabla V\right) \notag\\
=& -\left( 1- \frac{\nabla f\nabla f}{|\nabla f|^2}\right)\cdot\nabla V - \frac{\nabla f}{|\nabla f|^2}\frac{1}{m}\mb p\cdot\nabla\nabla f\cdot\mb p.
\end{align}
We recover Eq.~\eqref{eof_L} by Eq.~\eqref{p_is_mv}. This process of elimination is algebraically equivalent to using the Dirac bracket for deriving the equation of motion \cite{weinberg_2015}.
\section{Geometric meaning of the normal force}\label{geometry}
We are now ready to endow the normal force term in Eq.~\eqref{eof_L} with the meaning of curvature of the subspace $f(\mb x) = 0$. In addition to classical mechanics, we shall adopt technologies from undergraduate level differential geometry of curves and surfaces, comprehensively covered in Ref.\cite{spivak1970comprehensive, books/daglib/0090942, kobayashi}. A useful compilation of implicit function related formulae is documented in Ref.\cite{GOLDMAN2005632}. To be self-contained, we shall explain the essence of the mathematical formulae using basic calculus.
\subsection{d=2}
In $d=2$, the constraint $f(x,y) = 0$ defines an implicit curve in $\mathbb R^2$. It is convenient to introduce the parametric representation for the curve. $\mathbb R\to \mathbb R^2: \tau \to \mb x = (x, y)$\footnote{We preserve the notation $t$ for the physical time, and allow $\tau$ to have generic meanings, but it can also refer to $t$. }. Within a finite range of $\tau$, we can define the arc length $s = \int^{\tau}d\tau' |d\mb x/d\tau'|$. The unit tangent vector $\mb t =|d\mb x/d\tau|^{-1} d\mb x/d\tau$ and the unit normal vector $\mb n$ \footnote{The orientation is, according to the most popular convention, defined by the right-hand rule.} at a point on the curve obey the Frenet equations 
\begin{subequations}
 \begin{align}
 & \frac{d\mb t}{ds} =\kappa \mb n\\
 & \frac{d\mb n}{ds} = -\kappa \mb t.  
 \end{align}
\end{subequations}
$\kappa$ is the curvature of the curve. These equations entails that the planar curvature measures the extent of rotation of the unit tangent and unit normal vectors along the curve. They imply the following formulae for curvature computation
 \begin{align}
 \kappa = -\mathbf t\cdot\frac{d\mathbf n}{ds} = \mb n \cdot\frac{d\mb t}{ds}.
 \end{align}
In terms of the implicit function $f = 0$, the unit normal can be computed via its gradient $\mb n = \nabla f /{|\nabla f|} $.
\begin{align*}
\frac{d\mb n}{ds} = \frac{d\mb x}{ds}\cdot\nabla\frac{\nabla f}{|\nabla f|} = \frac{1}{|\nabla f|}\mb t\cdot\nabla\nabla f + \nabla f \times \mb t\cdot(\cdots).
\end{align*}
The details of the second trailing term is irrelevant because it is proportional to $\nabla f\propto \mb n$. Upon the projection onto $\mb t$,
\begin{align}
\kappa = - \frac{1}{|\nabla f|}\mb t\cdot\nabla\nabla f\cdot\mb t.
\end{align}
Physically, the velocity vector $\dot{\mb x}$ of the particle is the instantaneous speed $v$ times the unit tangent vector. Hence the normal force term in $d=2$ equals 
\begin{align}
\label{d=2_curvature}- \frac{\nabla f}{|\nabla f|^2}(\dot{\mb x}\cdot\nabla\nabla f\cdot\dot{\mb x}) = \mb n \kappa v^2.
\end{align}
The unit normal vector $\mb n$ is only defined up to a sign. $f(\mb x) = 0$ and $-f(\mb x) = 0$ have the same set of roots. Eq,~\eqref{d=2_curvature} is well-defined because $\kappa$ adjusts its sign correspondingly. For a simple closed curve, the {\it canonical} unit normal vector is defined inwardly pointing. 
\subsection{d=3}
The derivation in $d=3$ is a slight generalization of $d=2$ with the same spirit. In this case $f(\mb x) = f(x, y, z)$ defines an implicit surface. Again, we introduce, at least locally, a parametrization $\mathbb R^2\to \mathbb R^3: q^{\alpha} \to x^i$, where $\alpha = 1, 2$ and $i = 1, 2, 3$. There is no natural parametrization analogous to $s$ and the tangent vectors now belong to a tangent space spanned by ${\p \mb x}/{\p q^{\alpha}}$. The normal direction at $\mb x$ can be derived by taking the cross product 
\begin{align*}
\mb n \propto \frac{\p\mb x}{\p q^{1}}\times\frac{\p\mb x}{\p q^2}.
\end{align*} 
The geometric quantity that encodes the curvature in this set up is the second fundamental form
\begin{align}
\mathrm{I\!I} = -d\mb x\cdot d\mb n = -dq^{\alpha}\frac{\p\mb x}{\p q^{\alpha}}\cdot d\mb n.
\end{align}
The total derivative $d$ is subject to the variation on the surface. The second fundamental form measures the projection of the rotated normal vector $\mb n$ onto tangent space. To recall its implication, consider a curve passing through a point $\mb x_0$ on the surface, and its unit tangent is $\mb t_0$. The normal curvature $\kappa_n$ of this curve at $\mb x_0$ is then given by $\mathrm{I\!I}(\mb t_0, \mb t_0)$. We now show that the normal acceleration in this scenario corresponds to the second fundamental form evaluated at $\mb v$. Again using the implicit function, the unit normal vector can be expressed in terms of its gradient 
\begin{align*}
n_i = |\nabla_{\mb x} f|^{-1}\frac{\p f}{\p x^i} := g\frac{\p f}{\p x^i}. 
\end{align*}
Its total derivative on the surface simply is 
 \begin{align}
 &dn_i = dq^{\alpha}\frac{\p}{\p q^{\alpha}}\left(g \frac{\p f}{\p x^i}\right) \notag\\
 =& n_i d\ln g  + dq^{\alpha}\frac{\p x^j}{\p q^{\alpha}} g\frac{\p^2f}{\p x^i \p x^j}.
 \end{align}
 The tangent differential is simply $dx^j = dq^{\beta}\frac{\p x^j}{\p q^{\beta}}$. It is then clearly, 
\begin{align}
 & \mathrm{I\!I} (\mb v, \mb v) = -\frac{dq^{\alpha}}{dt} \frac{dq^{\beta}}{dt} \frac{\p x^j}{\p q^{\beta}}\frac{\p x^i}{\p q^{\alpha}}\frac{1}{|\nabla_{\mb x} f|}\frac{\p^2 f}{\p x^i\p x^j}\notag\\
 =& -\mb v\cdot \frac{\nabla_{\mb x} \nabla_{\mb x} f}{|\nabla_{\mb x} f|}\cdot\mb v = v^2\kappa_n.
\end{align}
The result echos Eq.~\eqref{d=2_curvature}. A more mathematically precise derivation can be found in Ref.\cite{gallot2004riemannian}. Consolidating the conclusions from both dimensions, we can rewrite Eq.~\eqref{eof_L} as
\begin{align}
\label{master_eom}m \ddot{\mb x} = \mb F  = \mb F_t + \mb F_n= - (1-\mb n\mb n)\cdot\nabla V + \mb n mv^2\kappa_n.
\end{align}
These results are sensible based on dimensional analysis because the ratio of Hessian to gradient is one over length regardless of the function $f$. One another physical example that helps endorse these results is the case where $V(\mb x) = 0$. Energy conservation implies the speed has to be a constant and therefore for the particle trajectory the physical time $t$ is proportional to the arc length $s$. The equation of motion reduces to 
\begin{align}
\frac{d^2\mb x}{ds^2} = \frac{d\mb t}{ds}= -\mb n (\mb t \cdot\nabla\nabla f\cdot\mb t).
\end{align}
The full Frenet-Serret formulae $d\mb t/ds = \kappa \mb n$ immediately identify the amplitude of the normal force with the curvature of the trajectory, or the normal curvature when this trajectory lies on a 2-dimensional surface.
\subsection{Examples}\label{examples}
Let us adopt the curvature perspective to review a couple of examples from basic mechanics.

\begin{figure}
 \includegraphics[width=1.0\linewidth]{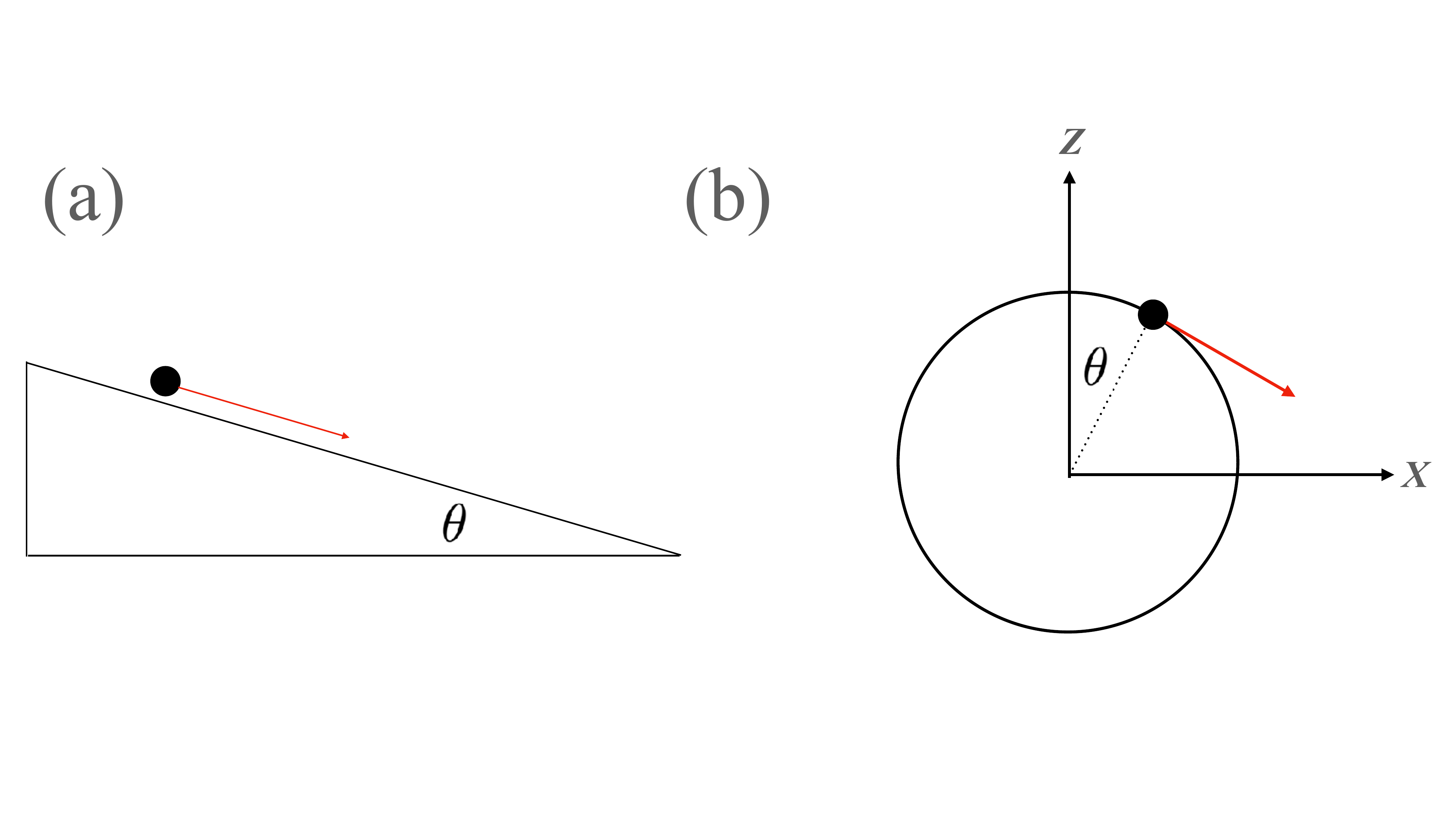}
 \caption{Illustrations of the geometry configurations for the example \ref{examplea} and example \ref{examplec}. }
 \label{fig0}
\end{figure}

\subsubsection{particle falling from a ramp}\label{examplea}
Suppose a particle of mass $m$ slides from rest on a ramp defined by $f(x, y) = \tan\theta x + y -y_0 = 0$. See panel (a) in Fig. \ref{fig0}. The potential is $V = mgy$. Elementary free-body diagram analysis quickly solves that the only nontrivial net force is parallel to the ramp with the magnitude $mg\sin\theta$. From a {\it constrained} perspective \footnote{Strictly speaking this constraint does not fall into our category in the introduction because the particle can leave the ramp from the above, but in this potential the conclusion remains the same.}, there is no net normal force because the ramp is flat. The projected gravitational force onto the unit normal $\mb n = (\sin\theta, \cos\theta)^T$ is
\begin{align}
m \ddot{\mb x}= & -[1 - \mb n\mb n]\cdot\nabla V\notag\\
= & -mg\left[ (0, 1) - (\sin\theta, \cos\theta) \cos\theta\right]^T \notag\\
= & mg\sin\theta[\cos\theta, -\sin\theta]^T.
\end{align}
\subsubsection{planar circular motion:} Suppose we launch a particle of mass $m$ into a (infinitesimally thin) smooth circular track defined by $f(x, y) = R^2 - x^2 - y^2 = 0.$  The gravity modeled by the potential $mgz$ does not come into play. The curvature is simply the reciprocal of the radius $R^{-1}$ and the inward normal vector is $\mb n = -R^{-1}(x, y)^T$. Consequently the normal force is the well-known $mv^2R^{-1}\mb n.$
\subsubsection{particle sliding from a sphere}\label{examplec}
This example can be regarded as a combination of the above two. This time the particle of mass $m$ starts sliding from the top of the sphere given by the constraint $R^2 - x^2-y^2 - z^2 = 0$. Without loss of generality, we choose this top to be $\mb x = (0, 0, R)$ and the corresponding gravitational potential is $V(\mb x) = mgz$. Due to rotational invariance, we can project this problem to the plane $y = 0$. Under this dimensional reduction the constraint becomes $R^2 - x^2 - z^2 = 0$. See panel (b) in Fig. \ref{fig0}. Hence, $\mb n = (-x/R, -z/R)^T:= -(\sin\theta, \cos\theta)^T$. The right-hand side of Eq.~\eqref{master_eom} are 
\begin{subequations}
\begin{align}
(1-\mb n\mb n)\cdot\nabla V & = mg\sin\theta (-\cos\theta, \sin\theta)^T\\
 \mb n\kappa v^2 & = \frac{mv^2}{R}(-\sin\theta, -\cos\theta)^T \notag\\
& = \frac{2mg(R-z)}{R}(-\sin\theta, -\cos\theta)^T,
\end{align}
\end{subequations}
where $mv^2$ is replaced using energy conservation. This problem has an interesting feature that $\lambda$ can vanish at finite $\theta$ by solving 
\begin{align}
\mb n\cdot\nabla V + \kappa mv^2 = 0\Rightarrow \cos\theta = \frac{2}{3}.
\end{align}
Post this value it becomes a non-holonomic problem.
\section{Geodesic equation}\label{geodesic}
In what follows, we will show, in $d=3$, when $V(\mb x) = 0$, the tangential component of the equation of motion is a geodesic equation on the constraint surface. This statement can be  motivated by the following observations. The first one is a gedanken experiment. Suppose we move a particle on a smooth sphere at constant speed away from other gravitational source. The trajectory of the particle would be a great circle if we are only exerting normal force. Another physical motivation is the action principle. let us introduce a local coordinate or parametrization for the constraint surface $\mathbb R^2\to \mathbb R^3: q^{\alpha}\to x^i$. We only assume this parametrization locally. Technically, if there is a global chart, we do not need to employ this machinery. In terms of the coordinate $q^{\alpha}$, $f(\mb x(q^{\alpha}))$ is fulfilled automatically and the action reduces to 
\begin{align}
\frac{1}{2}\int dt\, \dot q^{\alpha} \frac{\p\mb x}{\p q^{\alpha}}\cdot\frac{\p \mb x}{\p q^{\beta}}\dot{q}^{\beta} = \frac{1}{2}\int dt\, \mathrm{I}(\dot{q}, \dot{q}),
\end{align}
where $\mathrm{I}(\ ,\ )$ is the first fundamental form, or to most physicists, the metric. As a consequence, the action principle that minimizes the arc length square $g_{\alpha\beta}\dot q^{\alpha}\dot q^{\beta}$ mathematically also minimizes $\sqrt{g_{\alpha\beta}\dot q^{\alpha}\dot q^{\beta}}$ \cite{Nakahara:2003nw,carroll2003spacetime}. See also Sec.19 of Ref.\cite{arnold1989mathematical}. Mathematically, the equation of motion with $V=0$, $\frac{d^2\mb x}{ds^2} = \kappa_n\mb n$ suggest the geodesic vector of this curve vanishes everywhere, and thus the trajectory has to a geodesic \cite{books/daglib/0090942, kobayashi}. To show this explicitly, let us project the equation of motion~\eqref{eof_L} onto the tangent plane spanned by $\xi^{\alpha}\frac{\p\mb x}{\p q^{\alpha}}$
\begin{align}
\xi^{\alpha}\frac{\p\mb x}{\p q^{\alpha}}\cdot\ddot{\mb x} = 0.
\end{align}
Expanding this equation using $\frac{d}{dt} = \frac{dq^{\beta}}{dt}\frac{\p}{\p q^{\beta}}$, we have
\begin{align}
\label{geo1}\xi^{\alpha} \frac{\p\mb x}{\p q^{\alpha}}\cdot\frac{\p \mb x}{\p q^{\beta}}\ddot{q}^{\beta} + \xi^{\alpha}\frac{\p\mb x}{\p q^{\alpha}}\cdot\frac{\p^2\mb x}{\p q^{\beta}\p q^{\gamma}}\dot q^{\beta}\dot q^{\gamma} = 0.
\end{align}
The first term is $\xi^{\alpha}g_{\alpha\beta}\ddot{q}^{\beta}$ by identifying the first fundamental form as the metric tensor on the surface. To rewrite the second term, we can use the fact
\begin{align*}
\frac{\p g_{\beta\ga}}{\p q^{\alpha}} = \frac{\p}{\p q^{\alpha}}\frac{\p\mb x}{\p q^{\beta}}\cdot\frac{\p\mb x}{\p q^{\ga}} = \frac{\p^2\mb x}{\p q^{\alpha}\p q^{\beta}}\cdot \frac{\p \mb x}{\p q^{\ga}} + (\beta\leftrightarrow \ga)
\end{align*}
to phrase the standard Christoffel symbol in terms of the derivatives of position vectors $\mb x$:
\begin{align}
\Gamma^{\ga}_{\alpha\beta} &= \frac{1}{2}g^{\ga\delta}\left( \frac{\p}{\p q^{\alpha}}g_{\beta\delta} + \frac{\p}{\p q^{\beta}}g_{\alpha\delta} - \frac{\p}{\p q^{\delta}}g_{\alpha\beta}\right)\notag\\
\label{connection}& = g^{\ga\delta}\frac{\p \mb x}{\p q^{\delta}}\cdot \frac{\p^2\mb x}{\p q^{\alpha}\p q^{\beta}}.
\end{align}
Plugging this back to Eq.~\eqref{geo1}, we obtain 
\begin{align}
\label{geo2}\xi_{\alpha}\left( \ddot q^{\alpha} + \Gamma^{\alpha}_{\beta\ga}\dot q^{\beta}\dot q^{\ga}\right) = 0.
\end{align}
The sum in the parenthesis needs to vanish because $\xi$ is arbitrary and dummy, implying the geodesic equation. We note although the explicit dependence on $f(\mb x)$ is projected out of this derivation, it is required because the local chart shall obey $f(x^i(q^{\alpha})) = 0.$

When $V(\mb x) \neq 0$, in general the particle trajectory will not be a geodesic on the constraint surface. For instance, one can easily confirm that for a particle moving on a sphere an external tangent force lead to a deviation of its trajectory from a great circle. To make a slight connection with contemporary spacetime physics, we note that the trajectory, nevertheless, thought of as the spacetime geodesic from the extended metric in the weak-potential approximation \cite{carroll2003spacetime,schutz:2009}:
\begin{align}
\label{weak_field_g}ds^2 = -\left( 1 + 2\phi \right) dt^2 + \left( 1 - 2\phi\right) g_{\alpha\beta}dq^{\alpha}dq^{\beta},
\end{align}
where $\phi$ is assumed to be a time-independent function. To verify this, on the one hand, let us also project the $-\nabla V$ term in Eq.~\eqref{eof_L} onto the tangent space. 
\begin{align*}
- \xi^{\alpha}\frac{\p x^i}{\p q^{\alpha}}\frac{\p V}{\p x^i} = -\xi_{\alpha}g^{\alpha\beta}\frac{\p V}{\p q^{\beta}},
\end{align*}
adding a external force to Eq.~\eqref{geo2},
\begin{align}
\label{geo3}\ddot q^{\alpha} + \Gamma^{\alpha}_{\beta\ga}\dot q^{\beta}\dot q^{\ga} = - \frac{1}{m}g^{\alpha\beta}\frac{\p V}{\p q^{\beta}}.
\end{align}
On the other hand, the time-like spacetime geodesic equation from the spacetime metric Eq.~\eqref{weak_field_g} is \cite{carroll2003spacetime,schutz:2009}
\begin{align}
\label{geo4}& \frac{d^2q^{\alpha}}{d\lambda^2} + \Gamma^{\alpha}_{00}\left(\frac{dq^0}{d\lambda}\right)^2 \notag\\
+& 2\Gamma^{\alpha}_{0\beta}\frac{dq^0}{d\lambda}\frac{d q^{\beta}}{d\lambda} + \Gamma^{\alpha}_{\beta\gamma}\frac{d q^{\beta}}{d\lambda}\frac{dq^{\gamma}}{d\lambda} = 0,
\end{align}
where $\lambda$ is an affine parameter parametrizing the geodesic. First we note that $\Gamma_{0\beta}^{\alpha} = 0$ since there is neither $0\alpha$ metric element nor time-dependence in Eq.~\eqref{weak_field_g}. As for the rest, using~\eqref{connection},
\begin{subequations}
\begin{align}
& \Gamma^{\alpha}_{00} = (1-2\phi)^{-1}g^{\alpha\beta}\frac{\p\phi}{\p q^{\beta}}\approx  g^{\alpha\beta}\frac{\p \phi}{\p q^{\beta}}\\
& \Gamma^{\gamma}_{\alpha\beta} \approx \frac{1}{2}g^{\ga\delta}\left( \frac{\p g_{\beta\delta}}{\p q^{\alpha}} + \frac{\p g_{\alpha\delta} }{\p q^{\beta}}- \frac{\p g_{\alpha\beta}}{\p q^{\delta}}\right) + \mathcal O(\p\phi).
\end{align}
\end{subequations}
In the non-relativistic regime, the proper time agrees with time $t$ in the leading order, and we can replace the affine parameter $\lambda$ with $t$. Consequently, Eq.~\eqref{geo4} reduces to Eq.~\eqref{geo3} with the identification $\phi = V/m$.
\section{Kinematic implications from differential geometry}\label{applications}
The theses proved in this work can make kinematic predictions about the normal force or the normal acceleration of a constrained motion by virtue of theorems from differential geometry. Some assertions about global properties can be considered {\it topological} and seem feasible for experimental realization. We shall elaborate four instances in $d=2$ and one in $d=3$.
\subsection{d=2}
Consider the case in $d = 2$. The curvature at a given point on the curve equals the normal acceleration divided by the square of instantaneous speed. 
\begin{align}
\kappa  = \frac{\ddot{\mb x}\cdot\mb n}{v^2} = \frac{a_n}{v^2}.
\end{align} 
This relation results to several kinematic implications when the constraint curve $\mathcal C$ is closed owing to known global properties of closed planar curves.
\subsubsection{rotation index theorem} 
Using the rotation index from on the Gauss sphere, the arc-length integral of the normal acceleration over speed square equals to a multiple of $2\pi$ regardless of the potential $V(\mb x)$.
\begin{align}
\oint_{\mathcal C} ds \frac{\mb n\cdot\ddot{\mb x}}{2\pi v^2} = \oint_{\mathcal C} dt \frac{\mb n\cdot\ddot{\mb x}}{2\pi v} = \oint_{\mathcal C} \frac{ds}{2\pi}\kappa \in \mathbb Z.
\end{align}
In the second formula, we substitute the time differential $dt$ for the arc-length $ds$ by $v = ds/dt$.
\subsubsection{bound on total curvature}
As an extension of the above case, we can also compute the arc-length integral of the absolute value of the curvature. Fenchel's theorem \cite{Fenchel1951OnTD} gives a bound on this integral over a closed curve. 
\begin{align}
\oint_{\mathcal C} ds \left|\frac{\mb n\cdot\ddot{\mb x}}{2\pi v^2} \right| = \oint_{\mathcal C} ds \frac{|\kappa|}{2\pi}\geq 1.
\end{align}
The equality saturates when $\mathcal C$ is simple, i.e., that it only coincides with itself once at the start and the end point, and convex.
\subsubsection{4-vertex theorem}
Suppose $\mathcal C$ is simple and closed. There are at least 4 extrema of the curvature on $\mathcal C$. In terms of the first order condition, the derivative of curvature with respect to arc-length vanishes at those 4 points. Using the kinematic correspondence to the normal acceleration, we arrive at  
\begin{align}
\frac{d}{ds} \frac{\mb n\cdot\ddot{\mb x}}{v^2} = 0.
\end{align}
at these extrema.
\subsection{d=3}
In $d = 3$, on one hand, the normal acceleration over speed square equals the normal curvature. On the other hand, the normal curvature can be expressed in terms of two principal curvatures $\kappa_1$ and $\kappa_2$ thanks to Euler's formula:
\begin{align}
\frac{a_n}{v^2} = \kappa_n = \kappa_1\cos^2\theta + \kappa_2\sin^2\theta,
\end{align}
where $\theta$ is the angle spanned by the velocity vector $\dot{\mb x}$ and the first principal direction. This relation yields a bound on the ratio of normal acceleration over speed square by the local first principal curvature. Taking the average of over the orientation $\theta$ yields the mean curvature 
\begin{align}
\left< \frac{a_n}{v^2}\right>_{\theta} = \frac{1}{2}(\kappa_1 + \kappa_2) = H.
\end{align}
One immediate inference is that if a particle is constrained on a minimal surface, where $H= 0$ by definition, on any point of the surface, its directional average of the normal acceleration over speed square is zero. 
\section{Concluding remarks}\label{conclusion}
We endowed the normal force exerted by the holonomic constraints of codimension $=$1 with a precise geometric meaning of curvature. Based on this identification, we explored a few kinematic implications, some of which hold regardless of the presence of the external potential. These results, accompanied by either physical or mathematical motivations, are fairly intuitive, echoing what one can anticipation from extrapolating contemporary dynamics. 

It is worth emphasizing we did not simply rephrase the thought process of the early differential geometricians. In their shoes, the trajectory is the starting point, from which they developed the concept of curvature. From our perspective, the given geometric data is the constraint space, and we interpret the dynamical quantity, the normal force, in terms of these geometric data, when the motion is confined.

In the same vein, this work bridges the conceptual leap from non-relativistic dynamics in flat a spacetime, Euclidean or Minkowski one, to a general curved manifold, by conceiving the curved space being embedded in a higher dimensional one. Though we have acknowledged and embraced the technical advantage of a mathematical framework without the notion of exteriority, it can be arguably pedagogically beneficial to perceive more advanced ideology via this intermediate step stone. Lab realizations of curvature related properties should be reasonably simpler as the object of interest is embedded within the scope of experiments. 

Moving forward, though we focused on the plain vanilla model of classical mechanics, the derivation is generic and generalizable. For instance, if the particle is electrically charged with charge $q$, the equation of motion is merely modified by the replacement $-\nabla V \to q(\mb E + \dot{\mb x}\times\mb B)$, and the results should hold in the leading order before accounting for radioactive reactions. Relatedly, a more comprehensive inclusion of dissipation is worth exploring. It should attribute an external source term to the tangential force, yet conclusions drawn directly from the normal force is expected to be valid.

Another relevant extension in $d=3$ is the case where the constraint subspace is an intersection of multiple constraints. A simple scenario is given by imposing two constraints $f(\mb x) = 0$ and $g(\mb x) = 0$ so that the resulting feasible space is a spatial curve. For example the intersection of a sphere $f(\mb x) = R^2 - x^2 - y^2 - z^2 = 0$ and a plane $g(\mb x) = z = 0$ is the equator of the sphere. The torsion of the curve is anticipated by the Frenet-Serret formulae.

Lastly, we look forward to additional physical or mathematical applications utilizing the results from this manuscript, and expect our point of view can complement the existing physics curriculum concerning the treatise on classical mechanics and kinematics. 
\begin{acknowledgments}
W.-H. thanks professor Pei-Ming Ho and Ping-Zen Ong for their enlightening lectures on quantum mechanics and differential geometry respectively back in early 2010s at the National Taiwan University.
\end{acknowledgments}
\bibliography{citation}
\end{document}